\documentclass[twocolumn,floatfix]{revtex4-1}
\usepackage{amsmath,amsthm,amssymb}
\usepackage[usenames]{color}
\usepackage{colortbl}
\allowdisplaybreaks

\theoremstyle{plain}

\usepackage{graphicx}
\newif \ifLastSection \LastSectionfalse

\numberwithin{equation}{section}

\newcommand{\be}{\begin{equation}}
\newcommand{\ee}{\end{equation}}
\newcommand{\ba}{\begin{eqnarray}}
\newcommand{\ea}{\end{eqnarray}}
\newcommand{\baa}{\begin{eqnarray}}
\newcommand{\eaa}{\end{eqnarray}}
\newcommand{\ed}{\end{document}}

\newcommand{\re}[1]{(\ref{#1})}
\newcommand{\ci}[1]{\cite{#1}}

\begin{document}

\title {Burgers equation on networks:\\ Metric graph based approach}
\author{K.K.~Sabirov$^{1}$, Kh. Sh. Matyokubov$^{2}$, and D.U.~Matrasulov$^{3,4}$}
\affiliation{$^1$University of Tashkent for Applied Sciences, 1 Gavkhar street, 100149 Tashkent, Uzbekistan\\
$^2$Urgench State Pedagogical Institute, 1 Gurlan str., 220100, Urgench, Uzbekistan\\
$^3$Turin Polytechnic University in Tashkent, 17
Niyazov Str., 100095, Tashkent, Uzbekistan\\
$^{4}$Tashkent University of Architecture and Civil Engineering, Yangishahar Str. 9A, 100095 Tashkent, Uzbekistan}

\begin{abstract}
 
 We consider Burgers equation on metric graphs for simplest topologies such as star, loops, and tree graphs.
 Exact traveling wave solutions are obtained for the vertex boundary conditions providing mass conservation and continuity of the solution at the nodes.  Constraints for the nonlinearity coefficients ensuring integrability of the Burgers equation are derived.  Numerical treatment of the soliton dynamics and their transmission through the graph vertex  is presented.
\end{abstract}
\maketitle
\section{Introduction}

Burgers equation is an important evolution equation having broad range of practical application. 
Since its derivation in pioneering work \ci{Burgers} the Burgers equation has found numerous
applications (see, e.g. the Refs.\cite{Khokhlov}-\cite{Dhawana}, review articles \ci{Bonkile18,Chowdhury,Yakushkin}  and books \ci{Saichev,Whitham,Sachdev}, for review).
In particular, it can be used to describe nonlinear waves in fluids
 and gases \ci{Saichev,Yakushkin,Fickett}, plasma \ci{Plasma},  blood vessels
\ci{Blood}, traffic flow \ci{Traffic,Chowdhury} and  in astrophysics \ci{Astro}. 
Depending on initial and other conditions, the Burgers equation has different traveling wave solutions, which describe kink solitons, shock and rarefactive waves and other types of the wave phenomena. 
General solution of the Burgers equation can be obtained by using so-called Hopf-Cole transformation\ci{Hopf,Cole},
which allows to reduce it into the linear form . Among most interesting from the viewpoint of practical applications are the solitons solutions \ci{Saichev,Sachdev,Soliton1,Soliton2}. Different traveling wave solutions are studied on the basis of Lax pair approach (see, e.g., \ci{LP}). Enstropy growth in the Burgers equation was studied in \ci{DP2012}.
Self-adjointness and conservation laws have been studied in the Refs.\ci{Nail}.
In \ci{GF} the Green function approach to Burgers equation was developed.
Different rarefaction solutions  of the Burgers equation are treated in \ci{Shock}.
Systematic review of the literature on solution methods  of the Burgers equation and different solutions can be found in review articles \ci{Dhawana,Bonkile}.

Wave dynamics in branched systems such as networks and lattices
has attracted much attention recently \ci{Zarif}-\ci{DSGE}.
Wave phenomena in such system can be described
by nonlinear evolution equation on metric graphs. The latter is
a set of the bonds connected to each other at the vertices. The
connection rule is called topology of a graph. In such approach
the problem is reduced to solving nonlinear wave equations for
which the boundary conditions at the branching points (vertices)  are imposed
\ci{Zarif}-\cite{Our1}. These boundary conditions are usually
derived from the fundamental conservation laws, such as energy, current, momentum, mass
or  charge conservation \ci{Zarif,Our1,DNNLSE}.
Recently the nonlinear Schr\"odinger
equations on networks have been studied in different contexts
(see, e.g. \ci{Zarif}-\ci{DP2015}) . Sine-Gordon equation on
networks were also considered \ci{EPL,DSGE} in the
context of Josephson junction networks. In \cite{JM} the Fokker-Planck equation on metrioc graphs was considered in the context of Brownian motion in branched strctures

In this paper we study the  Burgers equation on metric graphs by
focusing on such traveling wave solutions as rarefaction and rarefaction solutions.
Burgers equation on metric graphs and fractals is considered in\ci{Hinz}. Numerical solution of the visdous Burgers equation on star graph is presented in \ci{Shukla}. 
The Burgers equation on networks is of importance for the study of blood flow in branched vessels, plasma in branched systems, traffic flow in branched roads, fluid and gas dynamics in networks.
Designing of networks and other branched systems having desired conductivity of fluid, gas or plasma flow is required by many technological problems.
This paper is organized as follows. In the next section we give brief description of the Burgers equation on a line. Section III presents formulation of the problem for metric star graph and some solutions of the problem. In section IV we consider inviscid Burgers equation on metric graphs. Wave dynamics in networks described in terms of the Burgers equation on metric graphs is presented in section V. Finally, section VI presents some concluding remarks.

\section{Burgers equation on a line}
Here, following the Ref.\ci{Pelinovskii}, we briefly recall Burgers equation on a line.

The Burgers equation on a line can be written as

\begin{equation}
    \frac{\partial u}{\partial t}+\epsilon u\frac{\partial u}{\partial y}=\frac{\partial^2u}{\partial y^2},\,y\in(-\infty;+\infty),\label{beonline1}
\end{equation}
where the wave function $u\in C^2(-\infty;+\infty)$.

A  soliton solution of Eq.\re{beonline1} on a real line can be written as \ci{Pelinovskii}
\begin{equation}
u(y,t;\epsilon,v)=v\left(1-tanh\left[\frac{\epsilon
v(y-\epsilon vt)}{2}\right]\right),\label{sol1}
\end{equation}
where $\epsilon$ is the nonlinear coefficient, $v$ is a positive or negative parameter. 
The solution \re{sol1} fulfills the following asymptotic conditions:
\begin{equation}
u\to0,\text{ at }y\to+\infty,\; {\rm for}\; \epsilon>0,\; v>0,  
\label{asymp1}
\end{equation}
and
\begin{equation}
\frac{\partial}{\partial y}u\to0,\text{ at
}y\to\pm\infty, \;  {\rm for}\; \epsilon>0, \; V>0.
\label{asymp3}
\end{equation}

An interesting solution is called the rarefaction solution, or  rarefaction solution which is given on a finite interval, $|y|\leq b$ and can be written as
\begin{equation}
    u(y,t)=\frac{1}{\epsilon(t+a)}\left[y-b\tanh\frac{by}{2(t+a)}\right],\label{sol2}
\end{equation}
where $a$ is a constant.

\begin{figure}[h]
\begin{minipage}[h]{0.9\linewidth}
\center{\includegraphics[width=1\linewidth]{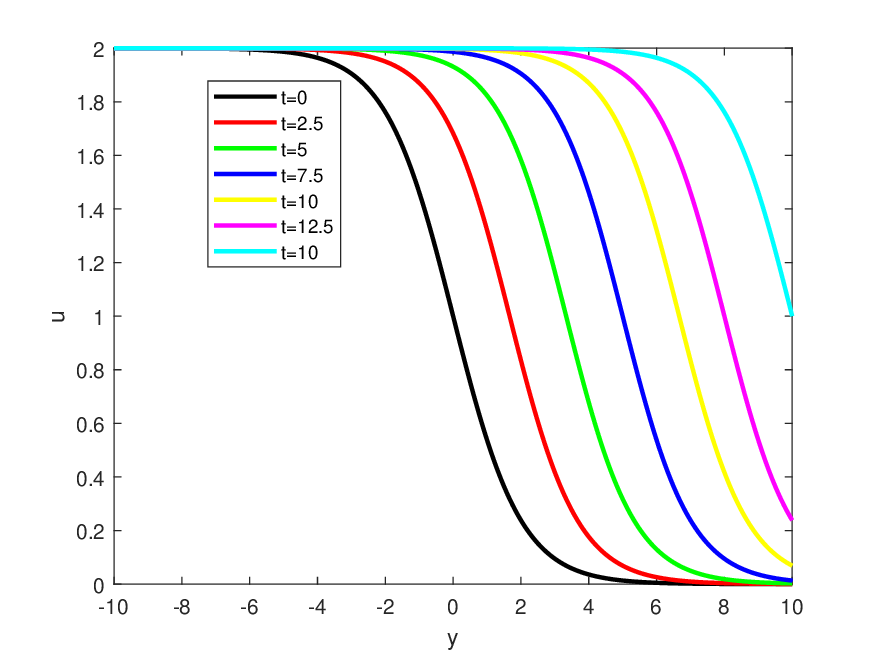}} a) \\
\end{minipage}
\vfill
\begin{minipage}[h]{0.9\linewidth}
\center{\includegraphics[width=1\linewidth]{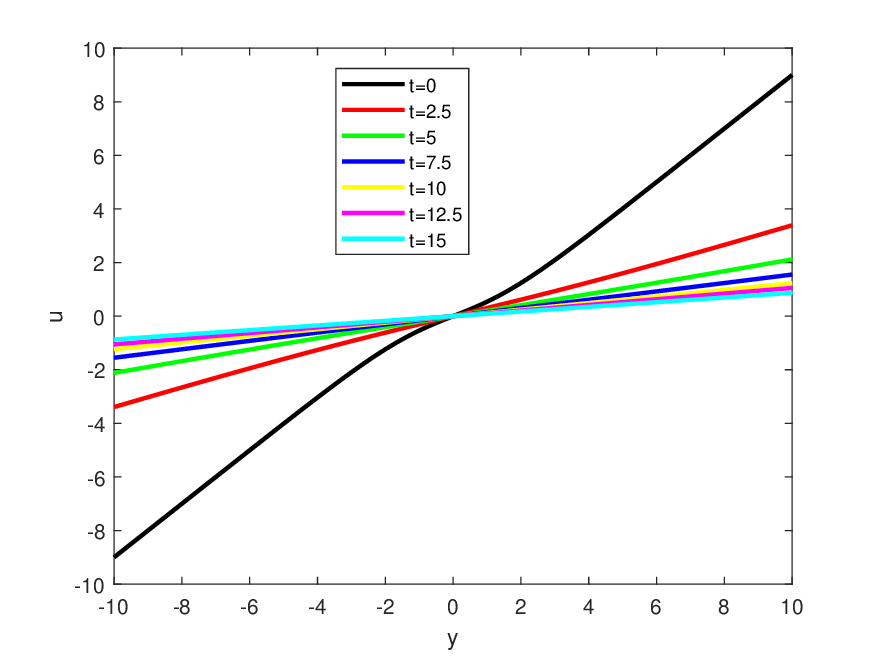}} b) \\
\end{minipage}
\caption{{\small  The soliton (a) and rarefaction (b) solutions of the Burger's equation on a line. 
\label{pic1}}}
\end{figure}

In Fig. 1., soliton and rarefaction solutions of the Burgers equation are plotted using Eqs.\re{sol1} and \re{sol2}, respectively. 

The special case of the Burgers equation is its  inviscid version, which does not contain second derivative over space variable and given by
\be
 \frac{\partial u}{\partial t}+\epsilon u\frac{\partial u}{\partial x}=0.
\label{beonline2} \ee

Solution of Eq.\re{be2} can written as
(\cite{Arora})
\begin{equation}
u(x,t)=\frac{1}{\epsilon}\frac{x}{1+t}.\label{sol3}
\end{equation}

\begin{figure}[h]
\center{\includegraphics[width=1\linewidth]{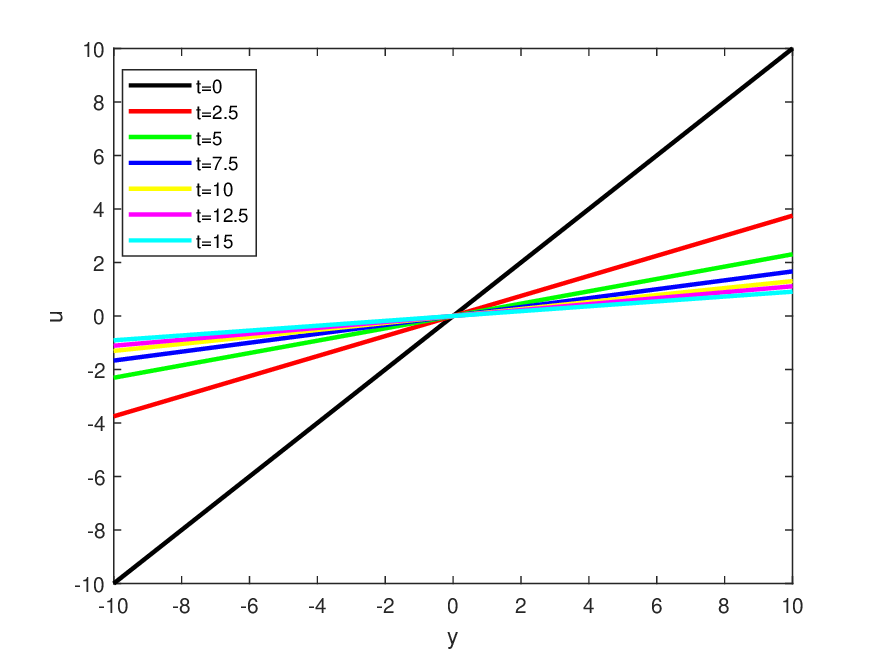}}
\caption{The plot of the solution of Eq. (ref{be2})}  \label{pic2}
\end{figure}
\section{ Burgers equation on metric star graphs: Boundary conditions and traveling wave solutions}

\subsection{Viscous Burgers equation}
Here we consider formulation and solution of the Burgers equation on a quasi-one-dimensional branched domain called metric graph. The graph itself present a system of wires connected to each other via some rule, which is called topology of a graph. When wires are assigned finite or semi-infinite length, the graphs is called metric graph.  To solve  an evolution  equation on a metric graph one needs to impose boundary conditions at the graph's branching points,
which are often called the vertex boundary conditions.
Such boundary conditions provide the connections of the
graph branches at the vertices.
\begin{figure}[h]
\center{\includegraphics[width=1\linewidth]{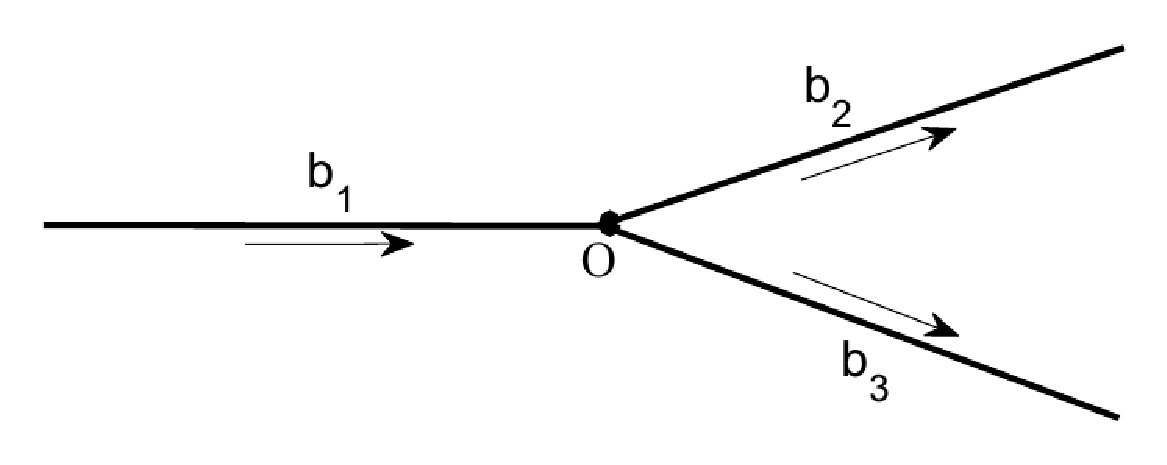}} \caption{Metric star graph}
  \label{pic3}
\end{figure}
We start from simplest graph, basic star graph  with three  semi-infinite
bonds, $b_1,\;b_2$ and $b_3$ (see, Fig. 2). The coordinates are defined as $x_1\in(-\infty,0]$ and
$x_{2,3}\in[0, \infty)$, with $0$ corresponding to the vertex point.

On each bond of the graph we can write  the  Burgers equation given by
\be
 \frac{\partial u_j}{\partial t}+\epsilon_ju_j\frac{\partial u_j}{\partial y}=\frac{\partial^2u_j}{\partial y^2}.
\label{be1}
\ee
where the functions, $u_j$ ( $j=1,2,3$ ) are assigned to
each bond of the graph. The set of the vertex boundary conditions for Eq. \re{be1} can be derived
from the fundamental conservation laws.  As the first set of the vertex boundary conditions we choose
continuity of the weights of the solution, which can be written as \be
|\epsilon_1u_{1}(0,t)|=|\epsilon_2u_{2}(0,t)|=|\epsilon_3u_{3}(0,t)|.
\label{vbc1} \ee

The second set of the vertex  boundary condition can be written as \be
\frac{1}{\epsilon_1}\left.\frac{\partial u_{1}}{\partial
y}\right|_{y=0}=\frac{1}{\epsilon_2}\left.\frac{\partial
u_{2}}{\partial
y}\right|_{y=0}+\frac{1}{\epsilon_3}\left.\frac{\partial
u_{3}}{\partial y}\right|_{y=0}, \label{vbc2} \ee
Below we show that the boundary conditions \re{vbc2} are consistent with the the mass conservation at the vertex.

The problem given by Eqs.\re{be1}, \re{vbc1} and \re{vbc2} present the problem of viscous Burgers equation on a metric star graph.

The viscous Burgers equation has broad variety of traveling wave solutions (see  \ci{Saichev,Whitham,Sachdev}, for review). Here we consider two types of such solutions, called soliton and  rarefaction solutions. 

Unlike inviscid Burgers equation, viscous one does not have infinitely many conservation laws.
The only conserved quantity for one-dimensional Burgers equation is the mass, which is given by
\begin{equation}
M=\underset{j=1}{\overset{3}{\sum}}\frac{1}{\epsilon_j}\underset{b_j}{\int}u_j(y,t)dy.\label{mass3}
\end{equation}
Theorem: In order for mass (\ref{mass3}) to be conserved, it is necessary and sufficient that the parameters $\epsilon_j$ fulfill the condition:
\begin{equation}
\frac{1}{\epsilon_1^2}=\frac{1}{\epsilon_2^2}+\frac{1}{\epsilon_3^2}.\label{sumrule1}
\end{equation}

Proof: We take time derivative and we use (\ref{be1}),
(\ref{vbc1})-(\ref{vbc2}) and
\begin{equation}
u_1\to0,\text{ at }y\to-\infty,\,u_{2,3}\to0,\,\text{at }y\to+\infty,  
\label{asymp3}
\end{equation}
and
\begin{equation}
\frac{\partial u_1}{\partial y}\to0,\text{at
}y\to-\infty,\,\frac{\partial u_{2,3}}{\partial y}\to0,\text{ at }y\to+\infty.
\label{asymp3}
\end{equation}
then we have
\begin{eqnarray}
&\frac{d}{dt}M=\underset{j=1}{\overset{3}{\sum}}\frac{1}{\epsilon_j}\underset{b_j}{\int}\frac{\partial}{\partial
t}u_j(y,t)dy=\nonumber\\
&=\underset{j=1}{\overset{3}{\sum}}\underset{b_j}{\int}\left(-u_j\frac{\partial
u_j}{\partial y}+\frac{1}{\epsilon_j}\frac{\partial^2}{\partial
y^2}u_j\right)dy=\nonumber\\
&=\frac{1}{2}\left(-u_1^2(0,t)+u_2^2(0,t)+u_3^2(0,t)\right)+\nonumber\\
&+\frac{1}{\epsilon_1}\left.\frac{\partial
u_{1}}{\partial
y}\right|_{y=0}-\frac{1}{\epsilon_2}\left.\frac{\partial
u_{2}}{\partial
y}\right|_{y=0}-\frac{1}{\epsilon_3}\left.\frac{\partial
u_{3}}{\partial
y}\right|_{y=0}=\nonumber\\
&=\frac{1}{2}\epsilon_1^2u_1^2(0,t)\left(-\frac{1}{\epsilon_1^2}+\frac{1}{\epsilon_2^2}+\frac{1}{\epsilon_3^2}\right)=0.\label{mass4}
\end{eqnarray}
The theorem is proved

Provided the parameters, $\epsilon_j$ fulfill the constraint \re{sumrule1},  the soliton solution the problem given by Eqs. \re{be1},  \re{vbc1} and \re{vbc2}, can be written as
\begin{equation}
u_{j}(y,t)=v_j\left(1-tanh\left[\frac{\epsilon_j
v_j(y+y_0-\epsilon_j v_jt)}{2}\right]\right)\,y\in
b_{j},\,j=1,2,3,\label{sol4}
\end{equation}
where  $v_{j}$ is a positive constant.
A solution of Burgers equation on metric star graph, called the rarefaction solution can be written as
\begin{equation}
u_j(y,t)=\frac{1}{\epsilon_j(t+a)}\left[y+y_0-b\tanh\frac{b(y+y_0)}{2(t+a)}\right].\label{sol5}
\end{equation}
Again, fulfilling the boundary conditions (\ref{vbc1})-(\ref{vbc2}) is ensured by the sum rule in (\ref{sumrule1}).

  \subsection{Extension for the tree graph}
The study presented in the previous section can be extended to the case of more complicated graphs, e.g. for tree graph,  
 presented in Fig. 4. On the each bond $b$ (corresponding
$b_1\sim(-\infty;0),\,b_{1i}\sim(0,L_{1i}),\,b_{1ij}\sim(0;+\infty),\,i=1,2,\,j=1,2,3$)
we have the Burgers equation (\ref{be1}) with vertices boundary
conditions \baa
&\epsilon_1u_1|_{y=0}=\epsilon_{1i}u_{1i}|_{y=0},\nonumber\\
&\epsilon_{1i}u_{1i}|_{y=L_{1i}}=\epsilon_{1ij}u_{1ij}|_{y=0},\label{vbc3}\\
&\frac{1}{\epsilon_1}\left.\frac{\partial u_{1}}{\partial
y}\right|_{y=0}=\frac{1}{\epsilon_{11}}\left.\frac{\partial
u_{11}}{\partial
y}\right|_{y=0}+\frac{1}{\epsilon_{12}}\left.\frac{\partial
u_{12}}{\partial
y}\right|_{y=0},\nonumber\\
&\frac{1}{\epsilon_{1i}}\left.\frac{\partial
u_{1i}}{\partial
y}\right|_{y=L_{1i}}=\underset{j=1}{\overset{3}{\sum}}\frac{1}{\epsilon_{1ij}}\left.\frac{\partial
u_{1ij}}{\partial y}\right|_{y=0},\label{vbc4} \eaa
where $i=1,2,\,j=1,2,3.$

\begin{figure}[h]
\includegraphics[width=1\linewidth]{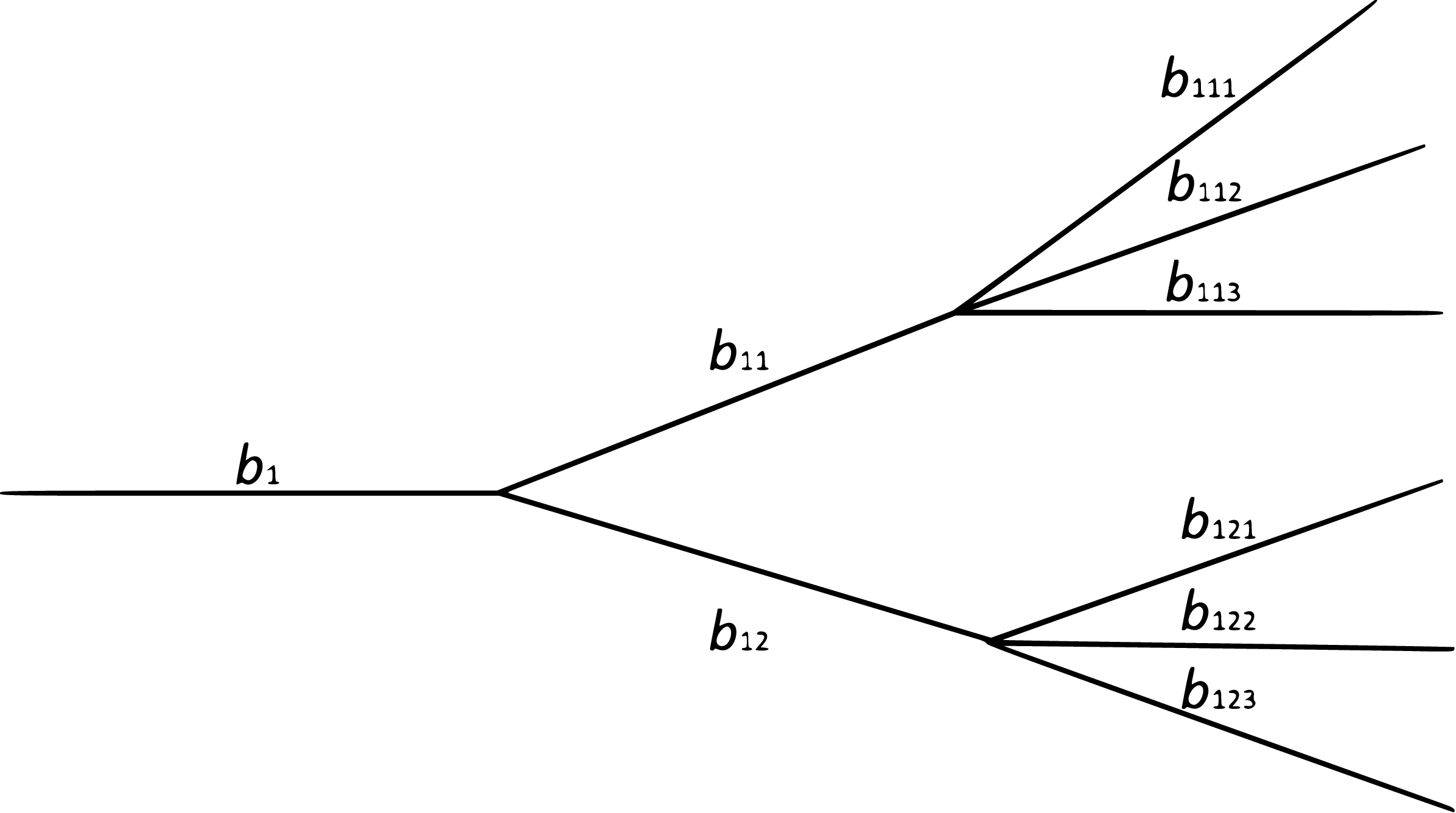} \caption{{Metric tree graph
  }}
  \label{pic7}
\end{figure}

On the each bond $b$ of the tree graph we have the following exact
solution 
\begin{equation}
u_{b}(y,t)=v_b\left(1-tanh\left[\frac{\epsilon_b
v_b(y+s_b-\epsilon_b v_bt)}{2}\right]\right),\label{sol7}
\end{equation}

where $b\in\{1,\,1i,\,1ij,\,i=1,2,\,j=1,2,3\}$,
$$s_{1}=s_{1i}=l,\,s_{1ij}=l+L_{1i},\,i=1,2,\,j=1,2,3.$$
Satisfying vertices boundary conditions we obtain the following
constrains \be
\frac{1}{\epsilon_1^2}=\frac{1}{\epsilon_{11}^2}+\frac{1}{\epsilon_{12}^2},\,\frac{1}{\epsilon_{1i}^2}=\underset{j=1}{\overset{3}{\sum}}\frac{1}{\epsilon_{1ij}^2},\,i=1,2.
\ee

\subsection{Inviscid Burgers equation}

Inviscid Burgers equation on star graph can be written (on each bond of the graph) as 
by \be
 \frac{\partial u_j}{\partial t}+\epsilon_ju_j\frac{\partial u_j}{\partial y}=0.
\label{be2} \ee where the wave functions $u_j$ are assigned to
each bond of the graph and   $j=1,2,3$ is the bond number.

The vertex boundary conditions for Eq.\re{be2} are given by Eqs.\re{vbc1} and \re{vbc2}.
Assuming that the sum rule in Eq. \re{sumrule1} is fulfilled the solution of the problem given by Eqs.
\re{be2},\re{vbc1} and \re{vbc2}, can be written as
\begin{equation}
u_j(y,t)=\frac{1}{\epsilon_j}\frac{y}{1+t}.\label{sol6}
\end{equation}

For inviscid Burgers equation thee mass is determined by Eq.\re{mass3}.
Similarly, to the above, one can prove that the mass conservation leads to the sum rule  \re{sumrule1}.

\begin{figure}[h]
\begin{minipage}[h]{1\linewidth}
\center{\includegraphics[width=1\linewidth]{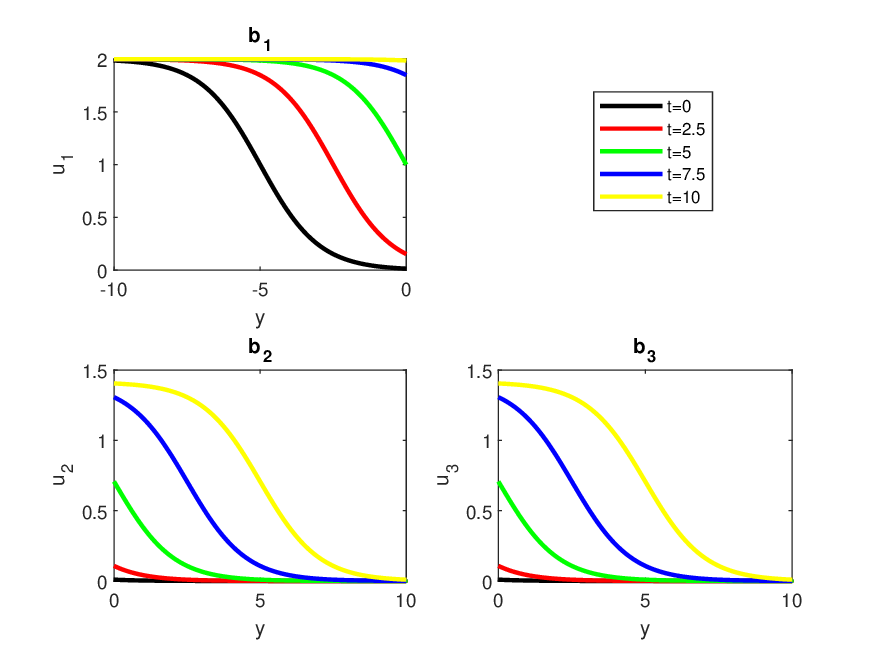}} a) \\
\end{minipage}
\vfill
\begin{minipage}[h]{1\linewidth}
\center{\includegraphics[width=1\linewidth]{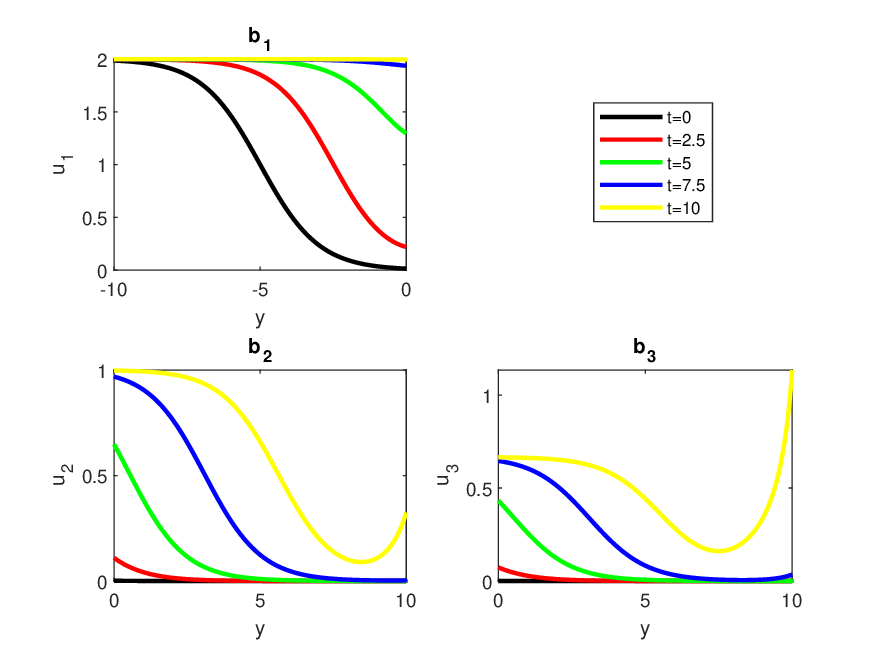}} b) \\
\end{minipage}
\caption{{\small The soliton solution of Eq.\re{be1} with the vertex boundary conditions, \re{vbc2} and \re{vbc1} are plotted for (a) integrable $\epsilon_1=1,\, \epsilon_2=\epsilon_3=\sqrt{2},\,v_1=1,\, v_2=v_3=\frac{1}{\sqrt{2}},\,y_0=5$ and (b) non-itegrable $\epsilon_1=1,\, \epsilon_2=2,\,\epsilon_3=3,\,v_1=1,\, v_2=\frac{1}{\sqrt{2}},\,v_3=\frac{1}{\sqrt{3}},\,y_0=5$
\label{pic4}}}
\end{figure}

\section{Wave dynamics and vertex transmission}

To explore wave dynamics in networks modeled by metric star graph, we consider the spatio-temporal evolution
of the traveling wave solutions for two regimes, for the regime when the constraints given by Eq.\re{sumrule1} are
fulfilled (integrable case) and broken (non-integrable case) by focusing on transmission of the waves through the branching points.

In Fig. \ref{pic4} rarefaction solution of Eq.\re{be1} for the vertex boundary conditions, \re{vbc2} and \re{vbc1} are plotted for (a) integrable
and (b) non-integrable cases. For both cases one can observe perfect(reflectionless) transmission of rarefactions through the vertex.
Such behavior is completely different than those observed in case of nonlinear Schrodinger \ci{Zarif}and sine-Gordon equations \ci{Our1} on metric graphs, which
exhibit reflectionless vertex transmission for integrable cases.

\begin{figure}[h]
\begin{minipage}[h]{1\linewidth}
\center{\includegraphics[width=1\linewidth]{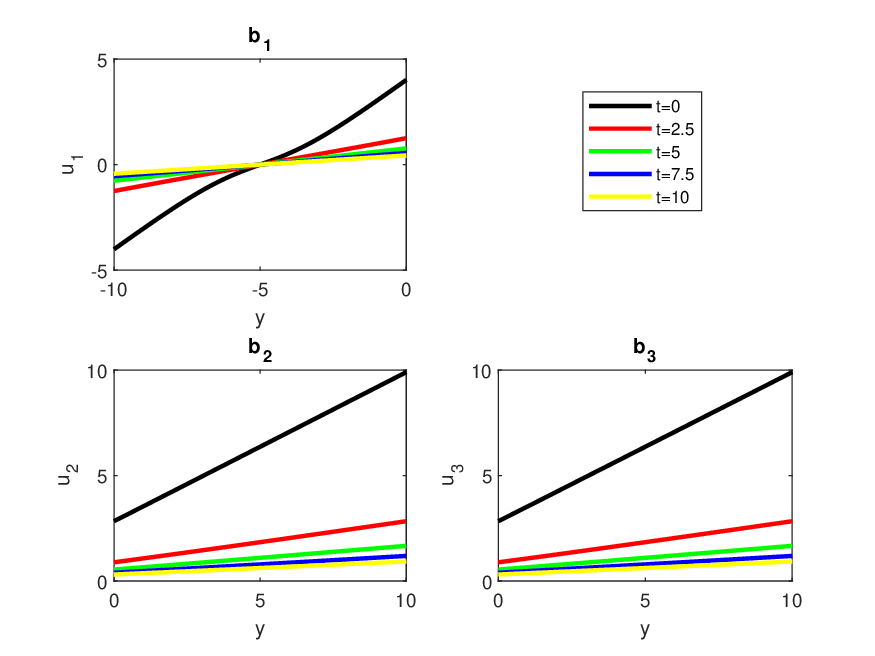}} a) \\
\end{minipage}
\vfill
\begin{minipage}[h]{1\linewidth}
\center{\includegraphics[width=1\linewidth]{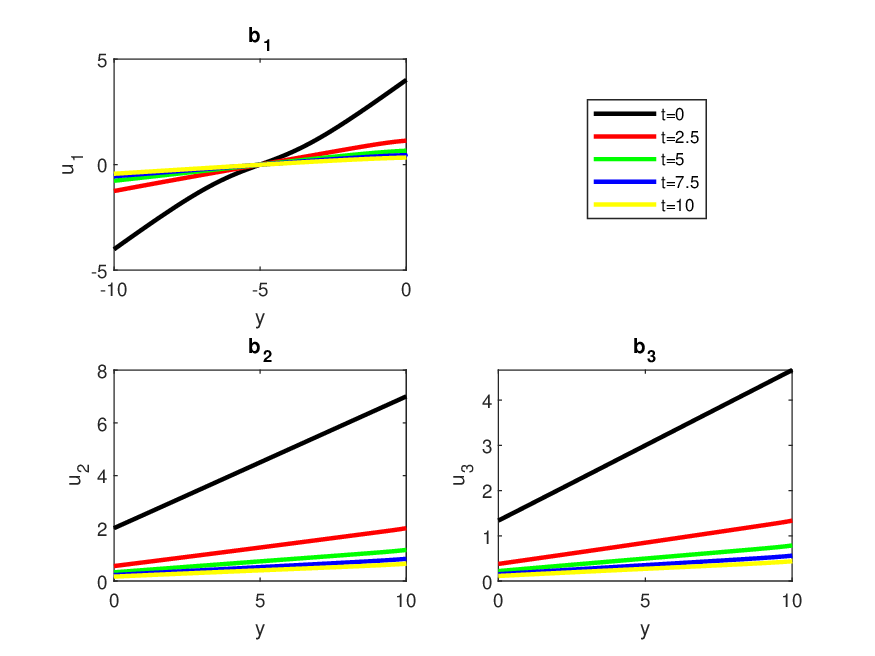}} b) \\
\end{minipage}
\caption{{\small The rarefaction solution of Eq.\re{be1} with the vertex boundary conditions, \re{vbc2} and \re{vbc1} are plotted for (a) integrable $\epsilon_1=1,\, \epsilon_2=\epsilon_3=\sqrt{2},\,a=1,\,b=1,\,y_0=5$
and (b) non-integrable cases $\epsilon_1=1,\, \epsilon_2=2,\,\epsilon_3=3,\,a=1,\,b=1,\,y_0=5$.
 \label{pic5}}}
\end{figure}

Fig. \ref{pic5} presents similar plots for the rarefaction solution of the Burgers equation on metric graphs
for (a) integrable and (b) non-integrable cases. The wave completely transmits through the vertex during some time period, in both cases.
However, after transmission it is localized near the vertex in second and third bonds. Finally, the solutions of inviscid Burgers equation given by Eq.\re{be2} and fulfilling the  vertex boundary conditions, \re{vbc2} and \re{vbc1} are plotted for  integrable 
and  non-integrable cases in Fig.7.

\begin{figure}[h]
\begin{minipage}[h]{1\linewidth}
\center{\includegraphics[width=1\linewidth]{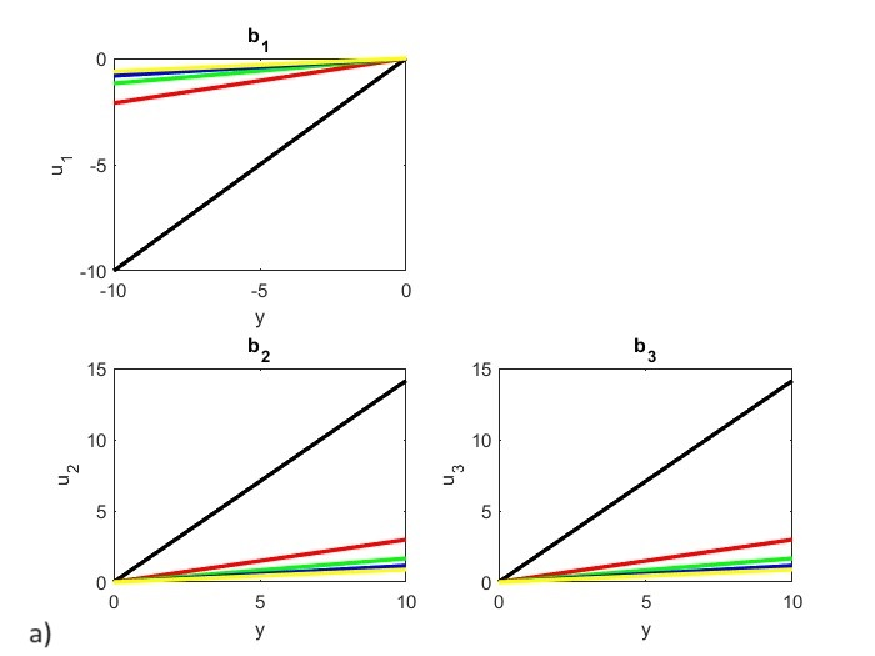}} a) \\
\end{minipage}
\vfill
\begin{minipage}[h]{1\linewidth}
\center{\includegraphics[width=1\linewidth]{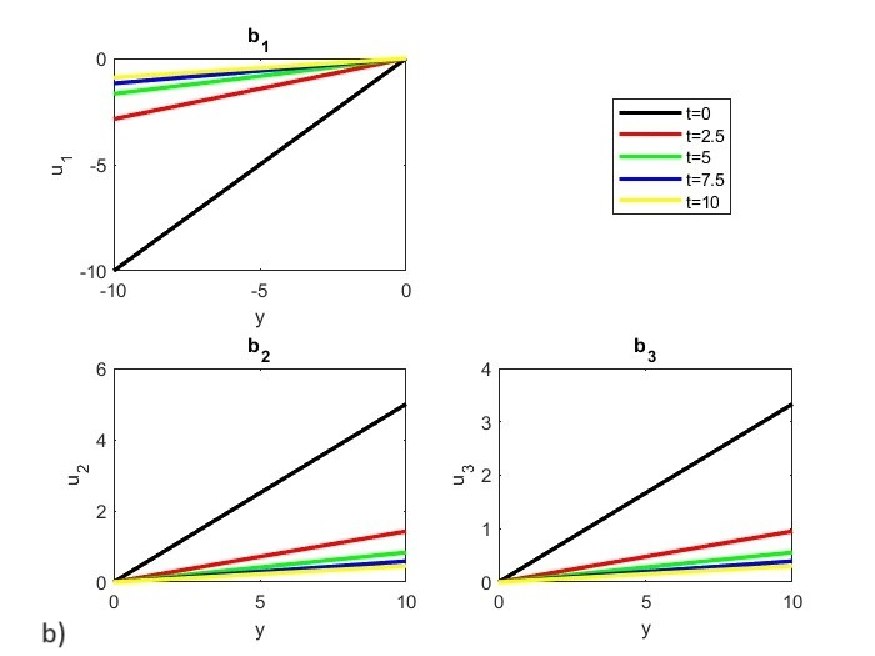}} b) \\
\end{minipage}
\caption{{\small The solution of inviscid Burgers equation given by Eq.\re{be2} with the vertex boundary conditions, \re{vbc2} and \re{vbc1} are plotted for (a) integrable $\epsilon_1=1,\, \epsilon_2=\epsilon_3=\sqrt{2},$
and (b) non-integrable cases $\epsilon_1=1,\, \epsilon_2=2,\,\epsilon_3=3$.
 \label{pic6}}}
\end{figure}
Both cases can be quite interesting from the viewpoint of practical applications in plasma, acoustics and traffic flow in branched structures.


\section{Conclusions and future challenges}

We studied several traveling wave solutions of the Burgers equation in networks by considering metric graphs. Rarefaction and kink soliton solutions on metric graphs are obtained for the vertex (matching) boundary conditions derived from mass conservation. Constraints that ensure integrability of the problem are expressed in terms of bond viscosity coefficients.

For non-integrable cases, the problem is solved numerically. An analysis of the transmission of the rarefaction solution through the graph vertex shows that the transmission is reflectionless in both integrable and non-integrable cases. For traveling waves described by rarefaction solutions, one can observe considerable scattering of the waves at the vertices and very slow transmission. 

Although the above treatment concerns very simple graph branching topologies, the approach can be directly extended to more complex network architectures and sizes. The above results may have potential practical applications in modeling tunable fluid, gas, plasma, and (road) traffic flow in branched structures. Especially, it can be attractive in the context of modeling blood vessels in the human body, where blood flows through branched domains. Another very interesting application involves vehicle traffic control, where models can be developed to minimize traffic jams and other complications.

\end{document}

\bibitem{Bonkile18} M. P. Bonkile, \textit{et.al}. Pramana J. Phys. {\bf} 90}69 (2018).